\documentclass[aps,prl]{revtex4}
\usepackage{graphicx}
\usepackage{hyperref}
\usepackage{amsfonts}
\usepackage{amsbsy}
\usepackage{amssymb}
\usepackage{xcolor}
\usepackage[normalem]{ulem}
 \usepackage{float}
 \usepackage{multirow}
\begin{document}

%\preprint{}

%Title of paper
\title{Time evolution of nonadditive entropies: The logistic map} 
\author{Constantino Tsallis$^{1,2,3,4}$}
\email{tsallis@cbpf.br}
\author{Ernesto P. Borges$^{5,2}$}
\email{ernesto@ufba.br}
\affiliation{ $^1$Centro Brasileiro de Pesquisas Fisicas\\
\mbox{Rua Xavier Sigaud 150, Rio de Janeiro 22290-180, Brazil} \\
 $^2$National Institute of Science and Technology for Complex Systems \\
\mbox{Rua Xavier Sigaud 150, Rio de Janeiro 22290-180, Brazil} \\
 $^3$ Santa Fe Institute, 1399 Hyde Park Road, Santa Fe, 
 New Mexico 87501, USA \\
 $^4$ Complexity Science Hub Vienna, Josefst\"adter Strasse 
 39, 1080 Vienna, Austria \\
 $^5$ Instituto de Fisica, Universidade Federal da Bahia, Salvador-BA 40170-115, Brazil
 }

\date{\today}

\begin{abstract}
Due to the second principle of thermodynamics, 
the time dependence of entropy for all kinds of systems 
under all kinds of physical circumstances always thrives interest. 
The logistic map $x_{t+1}=1-a x_t^2 \in [-1,1]\;(a\in [0,2])$ 
is neither large, since it has only one degree of freedom, 
nor closed, since it is dissipative. 
It exhibits, nevertheless, a peculiar time evolution of its natural entropy, 
which is the additive Boltzmann-Gibbs-Shannon one, 
$S_{BG}=-\sum_{i=1}^W p_i \ln p_i$, 
for all values of $a$ for which the Lyapunov exponent is positive, 
and the nonadditive one $S_q= \frac{1-\sum_{i=1}^W p_i^q}{q-1}$ 
with $q=0.2445\dots$ at the edge of chaos, 
where the Lyapunov exponent vanishes, 
$W$ being the number of windows of the phase space partition. 
We numerically show that, for increasing time, the phase-space-averaged entropy
overshoots above its stationary-state value in all cases. 
However, when $W\to\infty$, the overshooting gradually disappears 
for the most chaotic case ($a=2$), whereas, in remarkable contrast, 
it appears to monotonically diverge at the Feigenbaum point ($a=1.4011\dots$). 
Consequently, the stationary-state entropy value is achieved from {\it above}, 
instead of from {\it below}, as it could have been a priori expected. 
These results raise the question whether the usual requirements 
-- large, closed, and for generic initial conditions -- 
for the second principle validity might be necessary but not sufficient.
\end{abstract}
%\pacs{05.20.-y, 05.45.-a, 05.45.Ac}
% insert suggested keywords - APS authors don't need to do this
%\keywords{}
%\maketitle must follow title, authors, abstract, \pacs, and \keywords

\maketitle

\subsection{1 - Introduction}
Molecular physics may be seen as a more detailed description of solidly 
established laws of chemistry. 
Analogously, atomic physics may be seen as a more detailed description 
of solidly established laws of molecular physics. 
Nuclear physics, physics of elementary particles, evolve along the same lines. 
Epistemologically, it is tacitly required that, at each deeper and deeper 
description, the knowledge previously established on solid grounds 
is satisfactorily recovered at some adequate scale. 
Another paradigmatic example of the same path is general relativity which, 
in the $G \to 0$ limit, recovers special relativity, which in turn recovers, 
in the $c\to\infty$ limit, Newtonian mechanics. 
Analogously, quantum mechanics recovers, in the $\hbar \to 0$ limit, 
Newtonian mechanics. 
On the experimental world, optic microscopy was improved 
by electronic microscopy, in turn improved by scanning probe microscopy, 
and so on. 
The celebrated sentence 
``If I have seen further it is by standing on the shoulders of giants" 
that Isaac Newton included in his letter to Robert Hooke 
dramatically illustrates that same path. 
This is essentially how, along the years and centuries, 
the progress of sciences proceeds along the footprints of what was 
previously established on reliable bases.

In the realm of statistical mechanics, where the concept of coarse graining, 
hence of changements of scales, plays a foundational role, 
a similar path is being followed since its formulation in the XIXth century, 
and also along the last three-four decades. 
The pioneering works of Boltzmann and Gibbs 
\cite{Boltzmann1872, Boltzmann1877,Gibbs1901} 
established, upon undeniably solid bases, a magnificent theory which is 
structurally associated with the Boltzmann-Gibbs (BG) entropic functional 
\begin{equation}
S_{BG}=-k\sum_{i=1}^W p_i \ln p_i \;\;(\sum_{i=1}^W p_i=1)\,,
\label{BGentropy}
\end{equation} 
and consistent expressions for continuous or quantum variables, 
$k$ being a conventional positive constant adopted once forever 
(in physics, $k$ is chosen to be the Boltzmann constant $k_B$; 
in information theory and computational sciences, 
it is frequently adopted $k=1$). 

In the simple case of equal probabilities, this entropic functional 
is given by $S_{BG}=k \ln W$. Eq. (\ref{BGentropy}) is generically 
{\it additive} \cite{Penrose1970}. 
Indeed, if $A$ and $B$ are two probabilistically independent systems 
(i.e., $p_{ij}^{A+B}= p_i^A p_j^B$), 
we straightforwardly verify that 
$S_{BG}(A+B)=S_{BG}(A)+S_{BG}(B)$. 
This celebrated entropic functional is consistent with thermodynamics 
for all systems whose $N$ elements are either independent or weakly interacting
in the sense that only basically local (in space/time) interactions 
are involved. 
For example, if we have equal probabilities and the system is such that 
the number of accessible microscopic configurations is given by 
$W(N) \propto \mu^N\; (\mu>1; \,N\to\infty)$, 
then $S_{BG}(N)$ is {\it extensive} as required by thermodynamics. 
Indeed $S_{BG}(N)=k\ln W(N) \sim k(\ln \mu)N$. 
But if the correlations are nonlocal in space/time, $S_{BG}$ may become 
thermodynamically inadequate. Such is the case of equal probabilities 
with say $W(N) \propto N^\nu \;(\nu>0; \,N\to\infty)$: 
it immediately follows $S_{BG}(N) \propto \ln N$, 
which violates thermodynamical extensivity. 
To satisfactorily approach cases such as this one, 
it was proposed in 1988 
\cite{Tsallis1988,TsallisMendesPlastino1998,GellMannTsallis2004,Tsallis2009} 
to build a more general statistical mechanics based 
on the {\it nonadditive} entropic functional
\begin{equation}
S_q\equiv k\frac{1-\sum_{i=1}^W p_i^q}{q-1}
   =k\sum_{i=1}^W p_i \ln_q \frac{1}{p_i} 
   = -k\sum_{i=1}^W p_i^q \ln_q p_i 
   = -k\sum_{i=1}^W p_i \ln_{2-q} p_i \;\;
   (q \in \mathbb{R}; S_1=S_{BG})\,,
\end{equation}
with 
$\ln_q z \equiv \frac{z^{1-q}-1}{1-q} \; (\ln_1 z=\ln z)$ 
and its inverse 
$e_q^z \equiv [1+(1-q)z]_{+}^{1/(1-q)}$; 
($e_1^z=e^z$; $[z]_{+}=z$ if $z>0$ and vanishes otherwise); 
for $q<0$, it is necessary to exclude from the sum the terms with vanishing 
$p_i$. 
We easily verify that equal probabilities yield $S_q=k\ln_q W$, 
and that generically we have
\begin{equation}
\frac{S_q(A+B)}{k}=\frac{S_q(A)}{k}+\frac{S_q(B)}{k}
                   +(1-q)\frac{S_q(A)}{k}\frac{S_q(B)}{k} \,,
\end{equation}
hence
\begin{equation}
S_q(A+B)=S_q(A)+S_q(B)+\frac{1-q}{k}S_q(A)S_q(B) \,.
\end{equation}
Consequently, in the $(1-q)/k \to 0$ limit, we recover the $S_{BG}$ additivity. 
For the anomalous class of systems mentioned above, 
namely $W(N) \propto N^\nu$, we obtain, $\forall \nu$,  
the {\it extensive} entropy $S_{1-1/\nu}(N)=k\ln_{1-1/\nu}W(N) \propto N$, 
as required by the Legendre structure of thermodynamics 
(see \cite{Tsallis2009,TsallisCirto2013,Tsallis2022} and references therein). 

At this point let us remind that a general entropic functional 
$S_G(\{p_i\})$ is defined as {\it trace-form} if it can  be written as 
$S_G=k\sum_{i=1}^W f(p_i) \equiv k\sum_{i=1}^W p_i\,g(p_i) $. 
A $G$-generalized logarithm can be defined as 
$g(p_i)\equiv \ln_G (1/p_i)$. 
Consequently $S_G= k \langle \sigma_i \rangle$, 
where $\sigma_i\equiv \ln_G (1/p_i)$ is the {\it surprise} 
\cite{Watanabe1969} or {\it unexpectedness} \cite{Barlow1990}; 
$\sigma(p)$ is assumed to monotonically increase, when $p_i$ decreases 
from 1 to 0, from 0 to its maximum value (which may be infinity). 
Only trace-form entropies can be written as the mean value of a surprise.

Moreover, an entropic functional $S(\{p_i\}; \{\eta\})$ is said 
{\it composable} if it satisfies, for two probabilistically independent systems 
$A$ and $B$, the property 
$\frac{S(A+B)}{k}=F(\frac{S(A)}{k}, \frac{S(B)}{k};\{\eta\})$, 
where $\{\eta\}$ is a set of fixed indices characterizing the functional 
(e.g., for $S_q$, it is $\{\eta\} \equiv q$ ); 
we use the notation   $\{ 0\}$ to indicate absence of any such index. 
$F(x,y;\{\eta\})$ is a smooth function of $(x,y)$ which depends on a 
(typically small) set of universal indices $\{\eta\}$ defined in such a way 
that $F(x,y; \{ 0\})=x+y$ ({\it additivity}), 
which corresponds to the BG entropy. 
Additionally, $F(x,y;\{\eta\})$ is assumed to satisfy 
$F(x,0;\{\eta\})=x$ ({\it null-composability}), 
$F(x,y;\{\eta\})=F(y,x;\{\eta\})$ ({\it symmetry}), 
$F(x,F(y,z;\{\eta\}); \{ \eta \})=  F(F(x,y;\{\eta\}),z;\{\eta\})$ 
({\it associativity}) 
(see details and thermodynamical motivation in \cite{Tsallis2009,Tsallis2022}).  

We specially focus here on $S_q$ because the Enciso-Tempesta theorem 
\cite{EncisoTempesta2017} proves that this entropic functional 
is the unique one which is simultaneously trace-form, composable, 
and contains $S_{BG}$ as a particular case.

In the present paper, we numerically exhibit the above mentioned 
epistemological path on the time evolution of $S_q$(t) associated with
the paradigmatic one-dimensional dissipative logistic map.  
Through appropriate scaling of successively finer partitions of the phase-space
into $W$ equal windows, we verify that, for both strong and weak chaos, 
properties such as the entropy production per unit time 
and related phenomenology are satisfied. 

The logistic map is defined as follows:
\begin{equation}
x_{t+1}=1-ax_t^2\;(x_t \in [-1,1]; \, a \in [0,2]; t=0,1,2,\cdots)\,.
\end{equation} 
Depending on the value of the external parameter $a$, 
the corresponding Lyapunov exponent $\lambda$ can be positive, negative or zero.
When $\lambda >0$ we say that the system is 
{\it strongly chaotic}: the simplest, and strongest, such case emerges for 
$a=2$, which implies $\lambda =\ln 2 >0$. 
When $\lambda=0$ and the corresponding value of $a$, noted $a_c$, is located 
at the accumulation point of successive bifurcations, we say that the system is
{\it weakly chaotic}. 
The most studied such points occur at the edge of chaos, more precisely, 
at the so-called Feigenbaum-Coullet-Tresser point, 
with $a_c=1.40115518909205\dots$.  
In all cases, if we start from initial conditions such that the entropy nearly 
vanishes at $t=0$, we observe that, for all values of $q$, 
$S_q$ tends to increase (not necessarily in a monotonic manner) 
as time increases. 
But it tends to increase {\it linearly} (thus providing a {\it finite} 
entropy production per unit time) only for an unique value of the index $q$. 
For $a=2$, the entropy which {\it linearly} increases with time, 
thus yielding a {\it finite} entropy production per unit time 
(satisfying the Pesin identity for the entropy production per unit time 
$K_{BG} \equiv \lim_{t\to\infty} S_{BG}(t)/t =\lambda$), 
is $S_{BG}$. 
In contrast, at the edge of chaos, the entropy which {\it linearly} increases 
with time is $S_q$ with $q=q_c \equiv 0.24448770134128\dots$. 
In fact, depending on the initial conditions, there are infinitely many 
such linearities 
(see \cite{LyraTsallis1998,BaldovinRobledo2004,MayoralRobledo2005,Robledo2006} 
 and references therein). 
This is why we present here the corresponding mean values 
over a natural set of initial conditions, similarly to what has been calculated 
in \cite{AnanosTsallis2004}.

\subsection{2 - Strong and weak chaos}
\begin{figure}[htb]
\centering
\includegraphics[scale=0.40,angle=0]{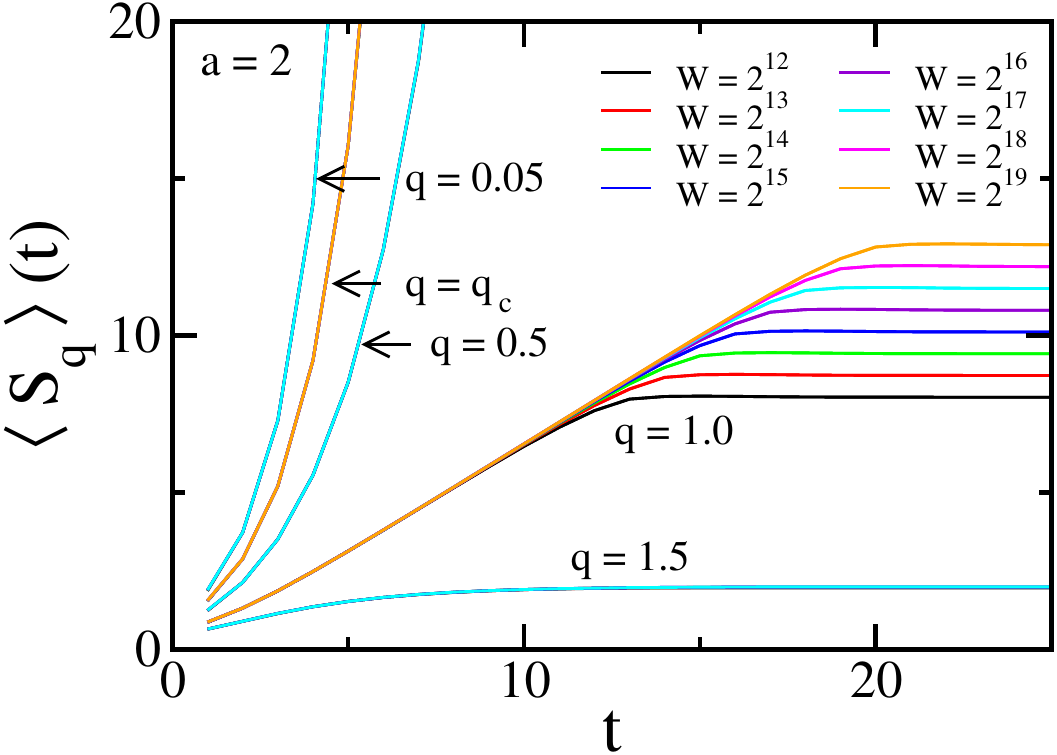}
\hspace{1.0cm}
\includegraphics[scale=0.40,angle=0]{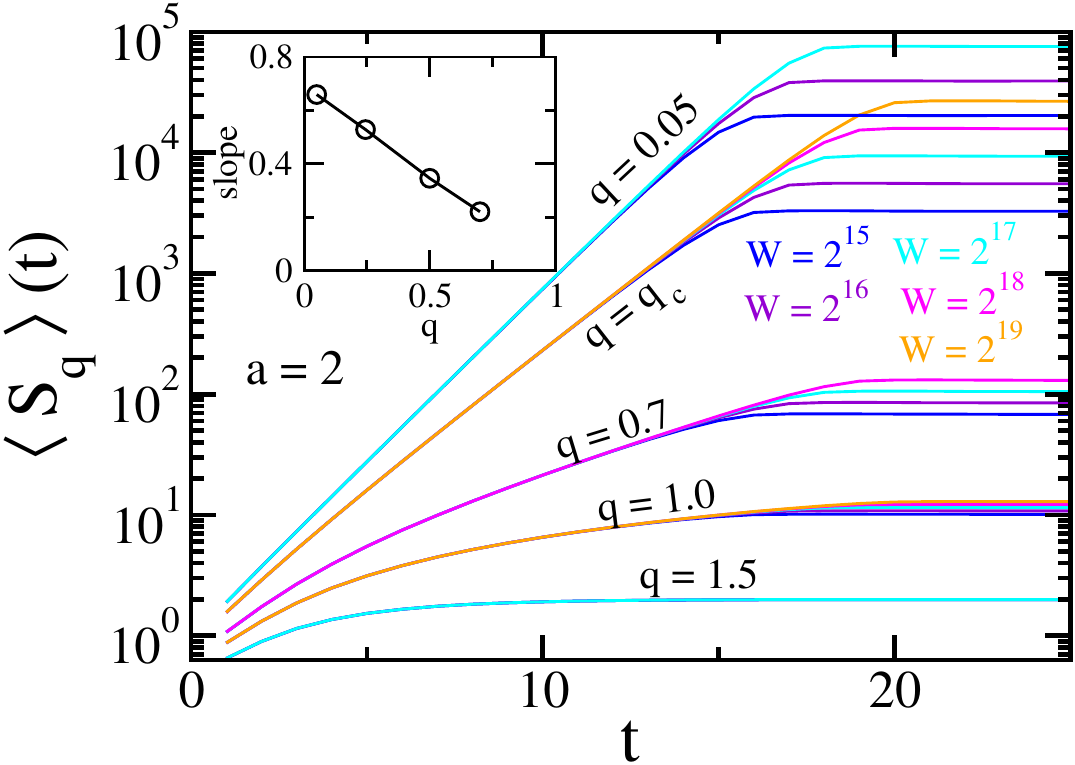}
\caption{\small (Color online) 
Time dependence of the average $\langle S_1 \rangle$ for $a=2$. 
{\it Left:} Linear-linear representation. 
The slope of the linear part equals $\lambda =\ln 2$, consistently with the 
Pesin identity. For all values of $q \ne 1$, several values of $W$ yield curves 
that are superimposed, and consequently can not be clearly identified. 
{\it Right:} Log-linear representation. 
We verify that, for $q<1$, a purely exponential behavior gradually emerges 
before saturation. 
Inset: The extrapolated slope for $q=0$ yields $0.69$.
}
 \label{fig1} 
\end{figure}

\begin{figure}[htb]
\centering
\includegraphics[scale=0.45,angle=0]{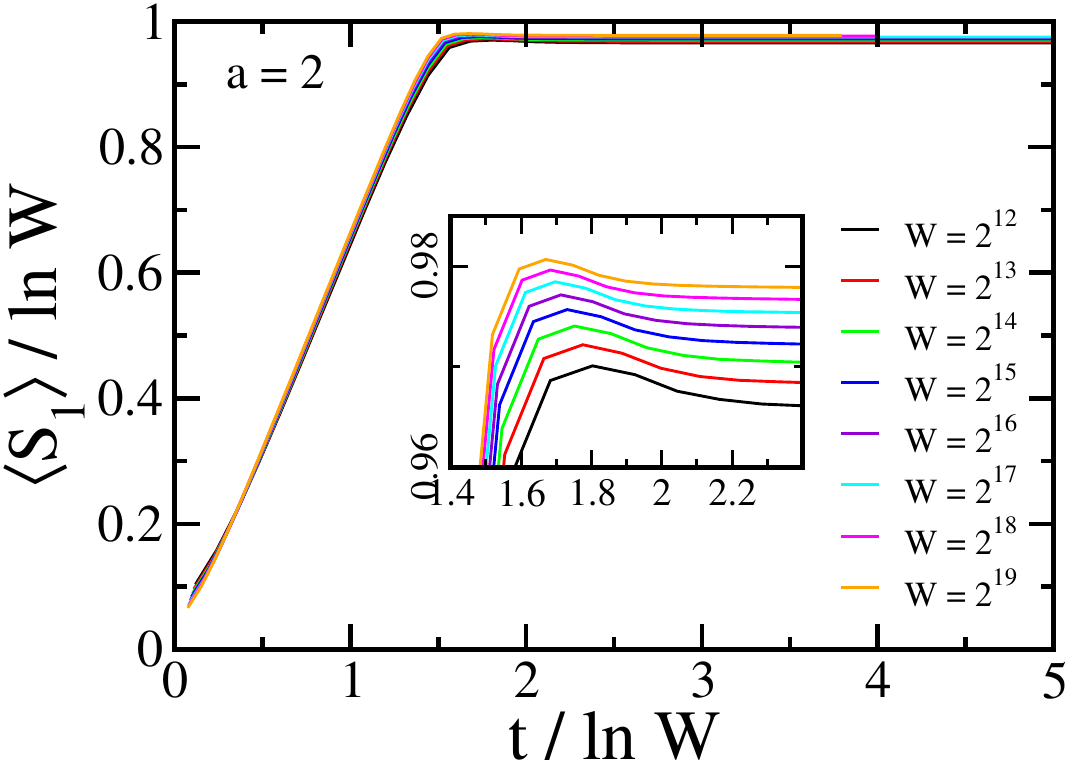}
\hspace{0.5cm}
\includegraphics[scale=0.45,angle=0]{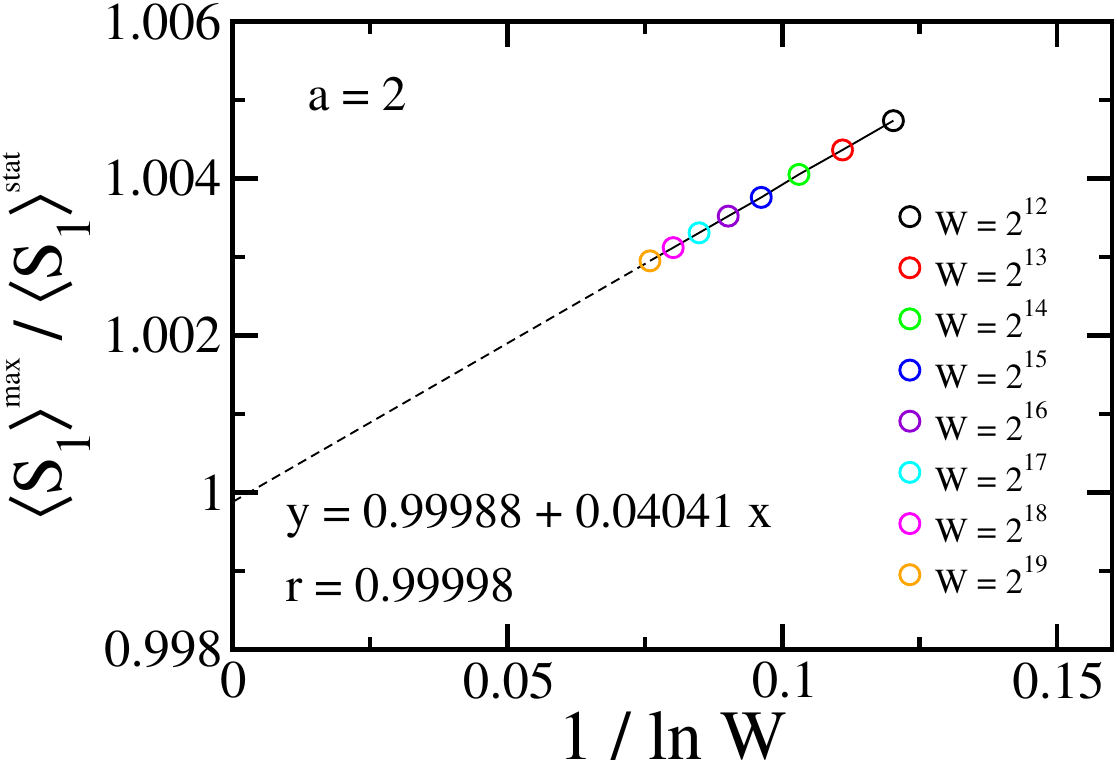}
\caption{\small (Color online) 
{\it Left:} Data collapse of the results indicated in Fig.\ \ref{fig1}. 
The maxima observed here are consistent with Fig.\ 1 of 
\protect\cite{BorgesTsallisAnanosOliveira2002}. 
{\it Right}: Extrapolation of the ratios. }
\label{fig2} 
\end{figure}

We partition the interval $x \in [-1,1]$ into $W$ equal little windows, 
and uniformly choose $M$ initial conditions,
typically $M=10 W$
(this number of initial conditions is sufficiently high for attaining 
 proper estimates of entropies. In fact, the relative error
 $\epsilon = \frac{1}{t_{\text{max}}} 
             \sum_{t=1}^{t_{\text{max}}} 
              \frac{|S_{q_c}(M=10W;t) - S_{q_c}(M=100W;t)|}{S_{q_c}(M=100W;t)}$
estimated with higher number of initial conditions, 
e.g.\ $M=100 W$ and $t_{\text{max}}=10000$ is $\epsilon < 2\times 10^{-3}$),
within one such interval (noted $j$). 
We denote $\{p_i\}\,(i=1,2,\cdots,W)$ the occupancy probabilities of all 
$W$ windows. 
At $t=0$, we have $p_{j}=1$ for the selected window, and $p_{i\ne j}=0$ for all the 
other $(W-1)$ windows. Consequently $S_q(t)$ satisfies $S_q(0)=0\,,\forall q$. 
For $a=2$, the only value of $q$ for which we have a linear growth 
while approaching saturation is $q=1$
(this procedure was initially proposed in 
\cite{LatoraBarangerRapisardaTsallis2000}).
We then repeat the operation for each one of the $W$ windows, 
and finally average the data for $S_q$: see Figs. \ref{fig1} and \ref{fig2}.


\begin{figure}[htb]
\centering
\includegraphics[scale=0.40,angle=0]{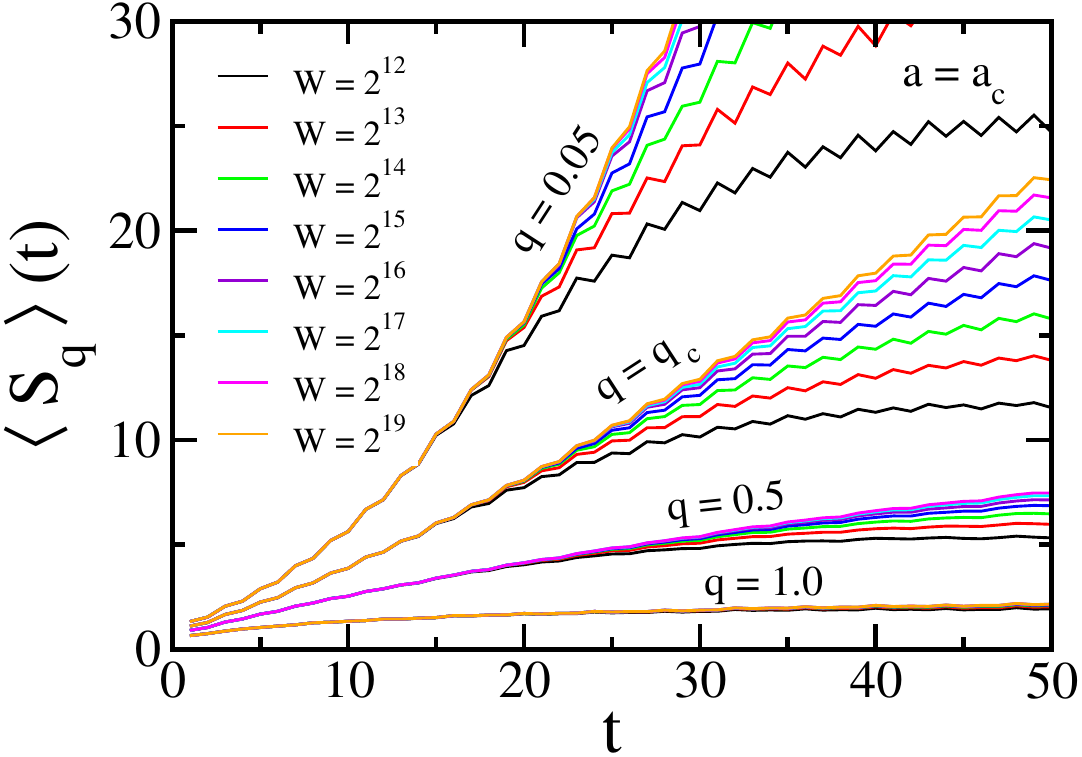}
\hspace{0.8cm}
\includegraphics[scale=0.40,angle=0]{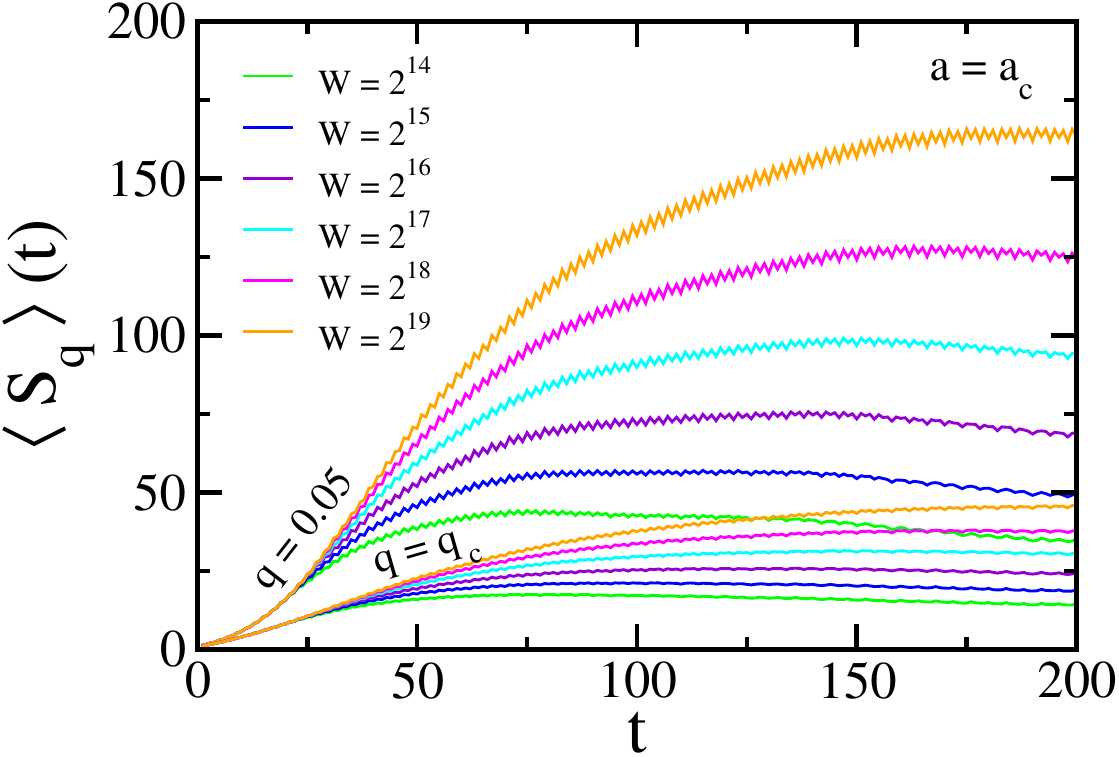}
\includegraphics[scale=0.43,angle=0]{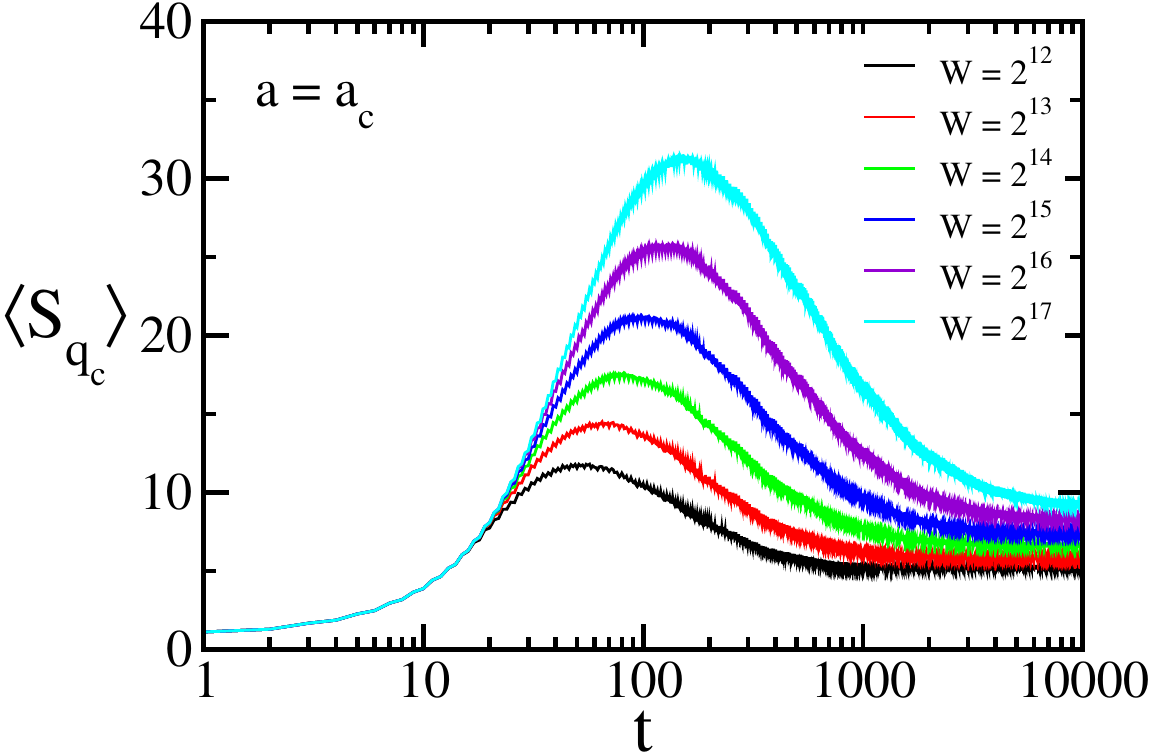}
\includegraphics[scale=0.43,angle=0]{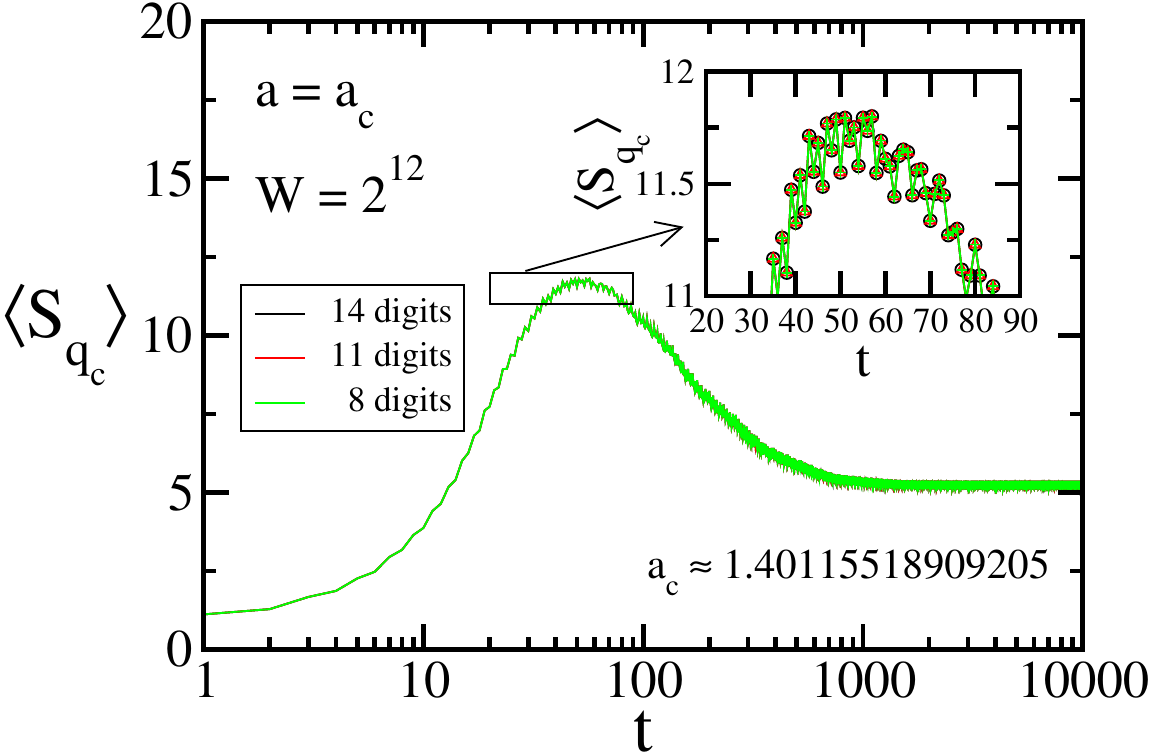}
\caption{\small (Color online) 
Time dependence of the average $\langle S_q \rangle$ for $a=a_c$. 
Strictly speaking, for the system to live 
(depending on the initial condition and assuming that all calculations 
 are done with infinite digits) 
on a multifractal, it is necessary to know $a_c$ with infinite digits, 
which is of course impossible. 
By varying the numerical precision that we have presently used for $a_c$, 
we have verified that this effect is numerically negligible 
for the number of digits within which our calculations have been done 
(see bottom-right plot).
}
\label{fig3}
\end{figure}

\begin{figure}[hb!]
\centering
\includegraphics[scale=0.43,angle=0]{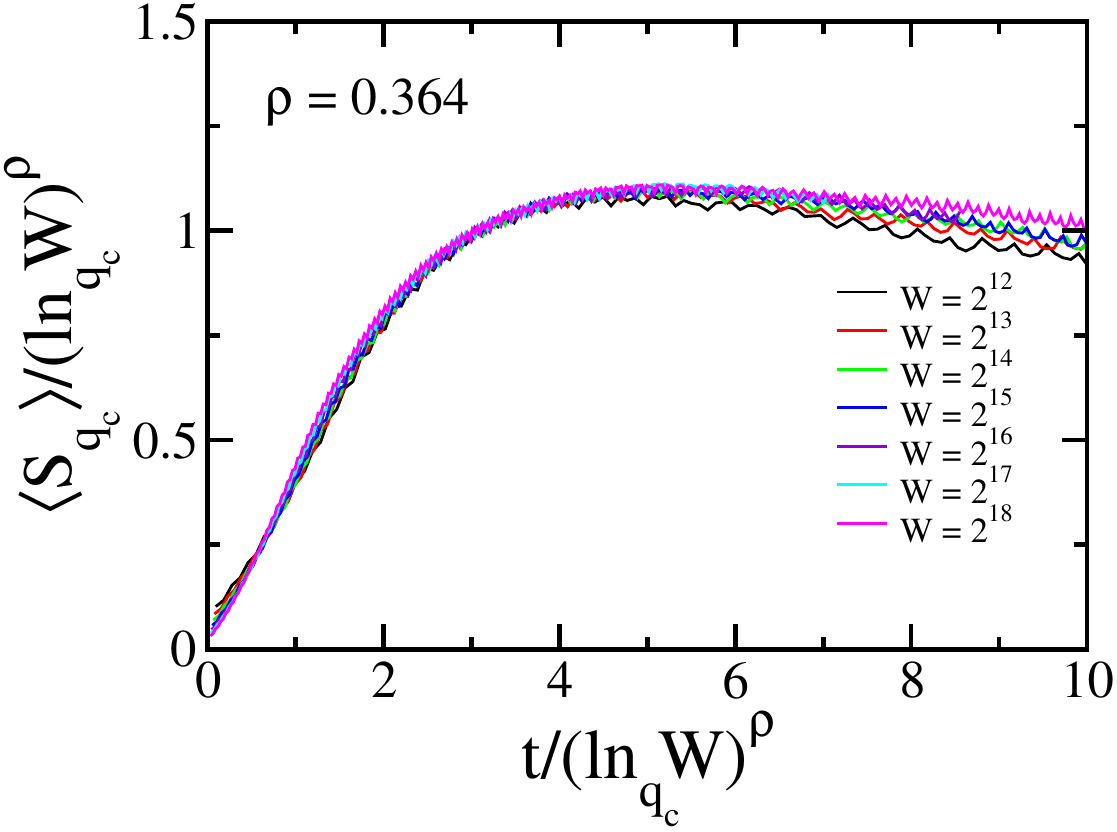}
\hspace{0.7cm}
\includegraphics[scale=0.43,angle=0]{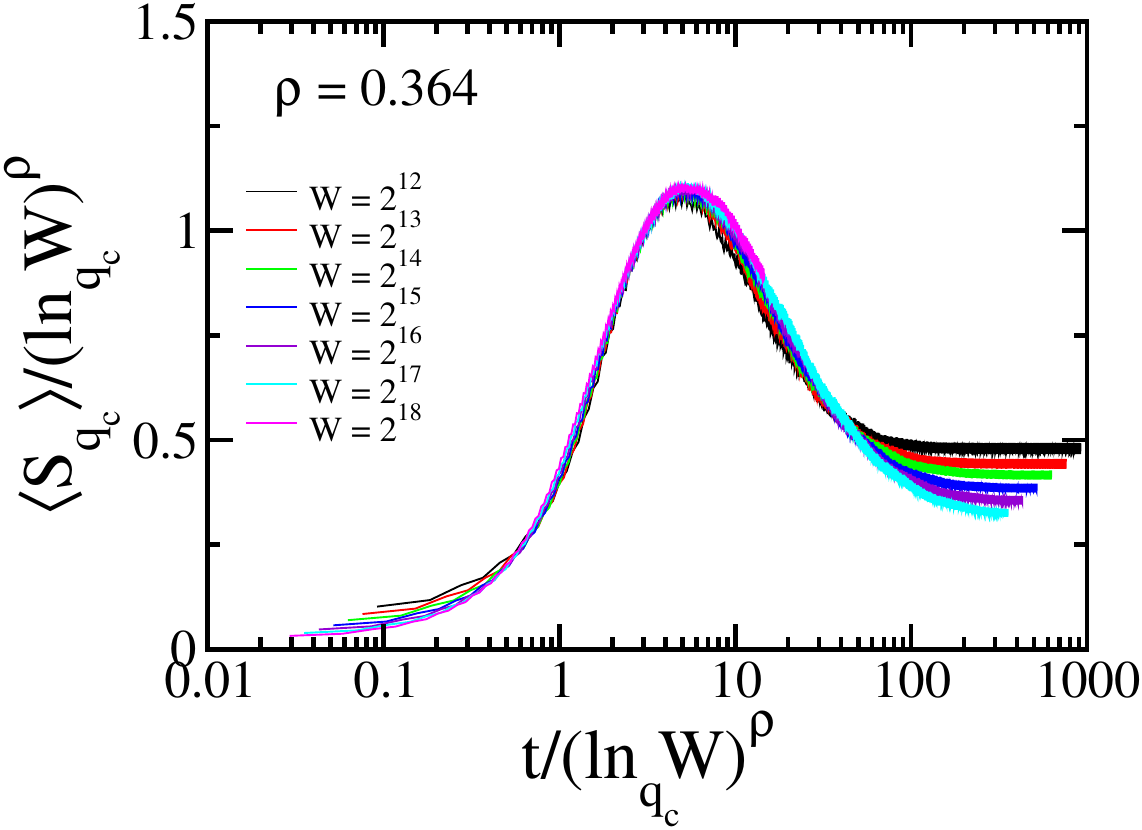}
\includegraphics[scale=0.43,angle=0]{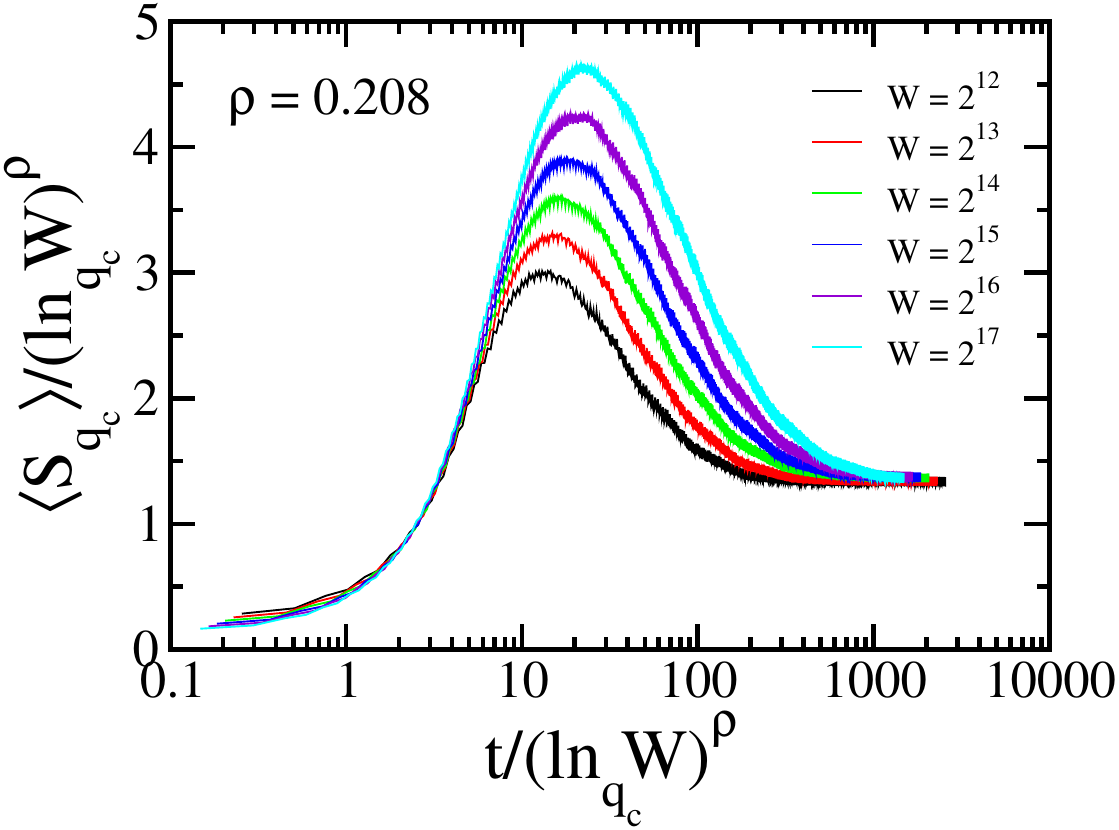}
\caption{\small (Color online) 
Data collapse of the results indicated in Fig.\ \ref{fig3}. 
Notice that, for $q<1$, $\rho>0$ and $W \gg 1$, 
$(\ln_q W)^\rho \sim [W^{1-q}/(1-q)]^ \rho \propto W^{(1-q)\rho}$, 
which reveals the (multi) fractal origin of $\rho$ 
(in the sense that $\rho$ is not an integer number). 
For $q \ne q_c$, there is no value of $\rho$ for which all the $a=a_c$ data 
can be collapsed through a single scaling with $(\ln_q W)^\rho$. 
{\it Top left:} Linear-linear representation aiming the collapse, 
for increasing $W$, of the maxima; 
the slope at first inflection point, located at $0.97 \pm 0.02$, 
is $0.47 \pm 0.01$; 
$\langle S_{q_c}\rangle/(\ln_{q_c}W)^\rho$ attains its maximal value 
$1.1 \pm 0.01$ at $t/(\ln_{q_c} W)^\rho = 5.3 \pm 0.3$, 
with $\rho = 0.364 \pm 0.003$.
{\it Top right:} The same as in Top-left but in linear-log representation. 
{\it Bottom:} Linear-log representation aiming the collapse, 
for increasing $W$, of the stationary-state heights; 
this collapse occurs at height $=1.35 \pm 0.04$ with $\rho=0.208 \pm 0.002$.
}
\label{fig4}
\end{figure}

\begin{figure}[ht!]
\centering
\includegraphics[scale=0.42,angle=0]{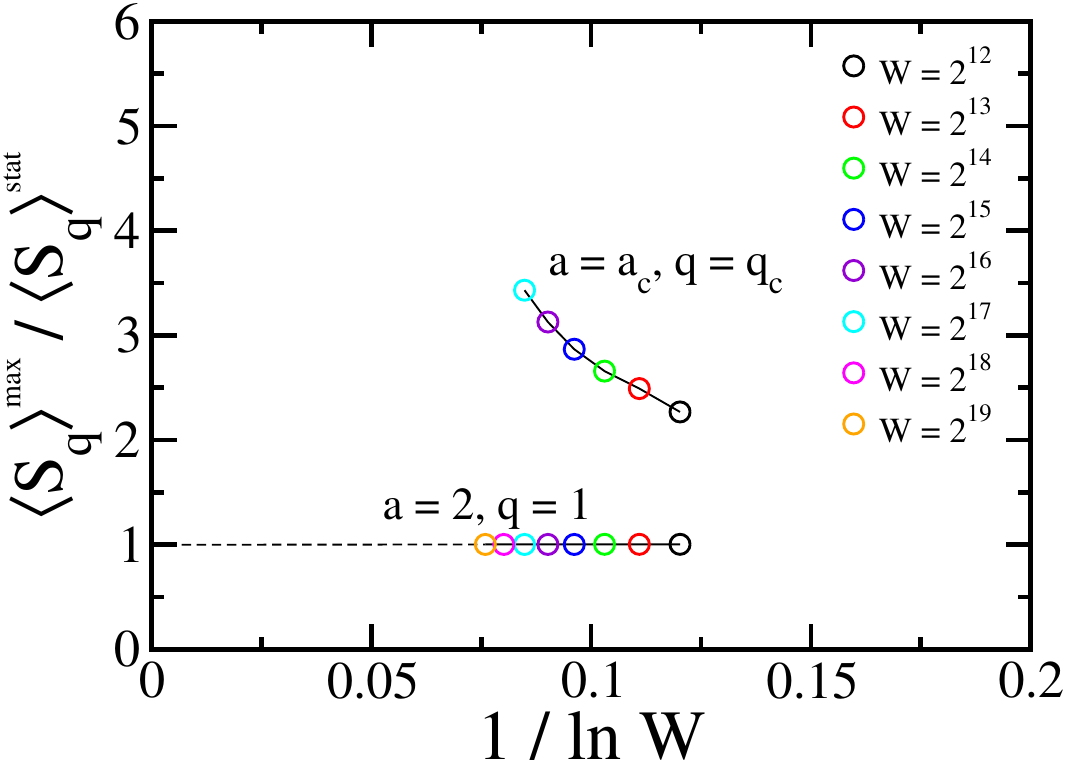}
\includegraphics[scale=0.41,angle=0]{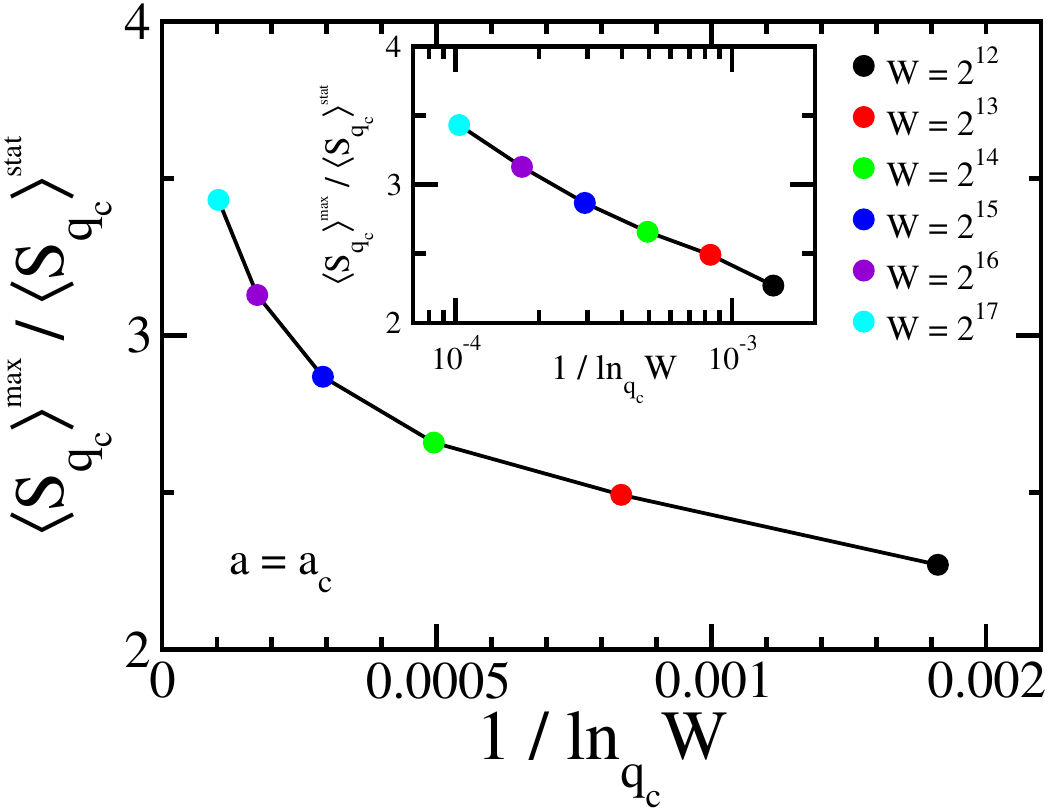}
\caption{\small (Color online) 
$W\to\infty$ extrapolations of the maxima. 
{\it Left:} Abscissa scaled with $\ln W$.   
{\it Right:} Abscissa scaled with $\ln_{q_c} W$.  
The data suggest that 
$\lim_{W\to\infty}[\langle  S_q\rangle^{\text{max}} / \langle S_q\rangle^{\text{stat}}]$
equals unity for $(a,q)=(2,1)$ and diverges for $(a,q)=(a_c,q_c)$.  
}
\label{fig5}
\end{figure}

We apply the same numerical procedure described above
for $a=a_c$: see Figs. \ref{fig3}, \ref{fig4} and \ref{fig5}.

\begin{figure}[h!]
\centering
\includegraphics[scale=0.45,angle=0]{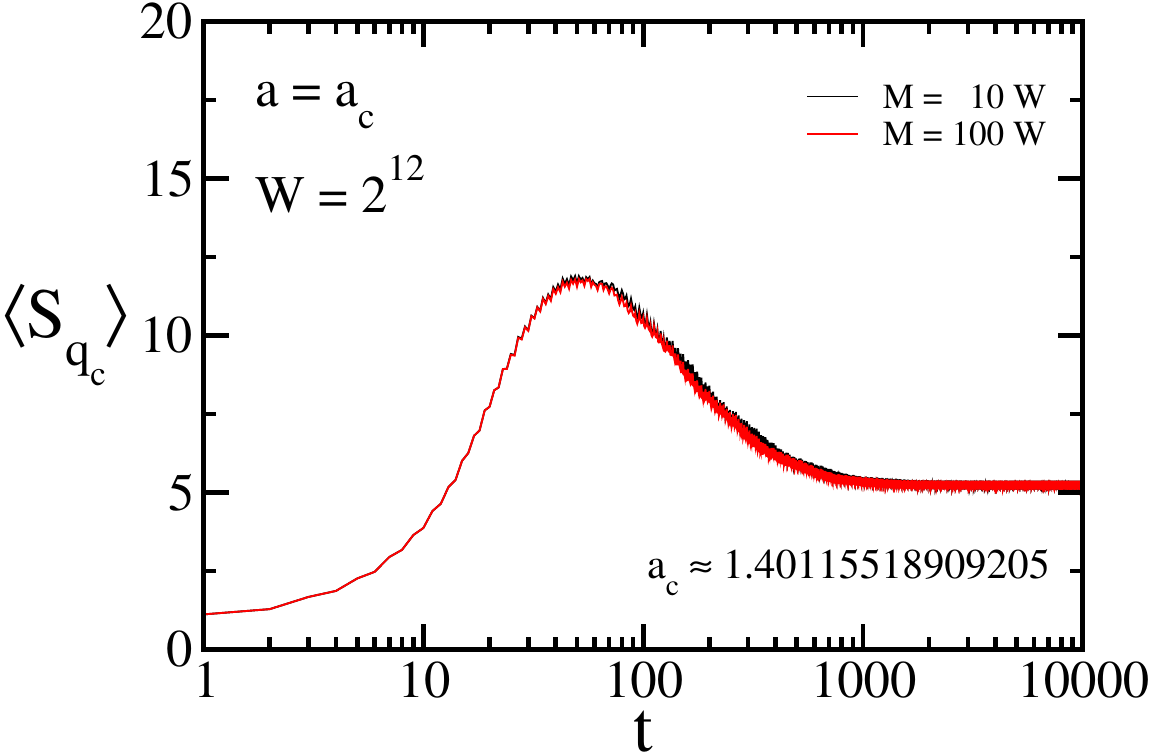}
\caption{\small (Color online) 
Influence of the number of initial conditions within each of the $W$ windows.
Instance with $W=2^{12}$ windows with two cases: $M = 10 W$ initial conditions
(black) and $M = 100 W$ initial conditions (red).
The similarity of the curves indicate that $M = 10 W$ is a sufficiently
large number of initial conditions for the purposes of the calculations.
}
\label{initialconditions}
\end{figure}

\begin{figure}[h!]
\centering
\includegraphics[scale=0.45,angle=0]{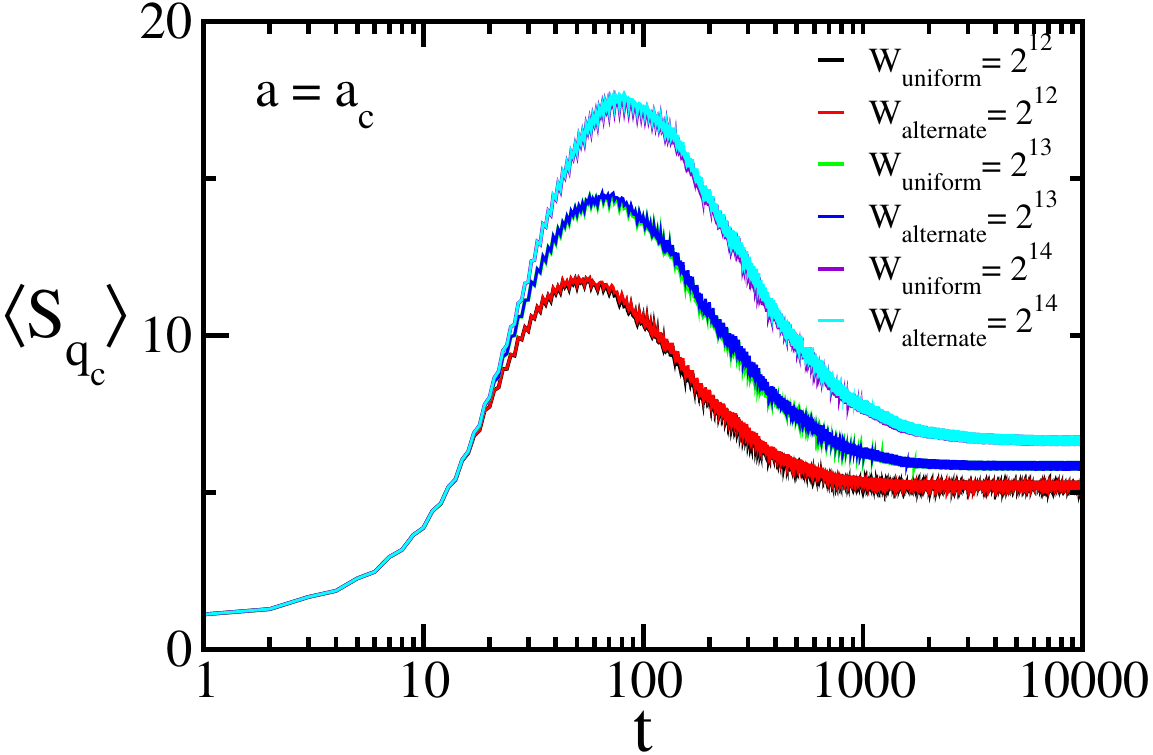}
\caption{\small (Color online) 
Influence of the phase-space occupation in the neighborhood of the 
multifractal attractor at $a=a_c$ (see \cite{RobledoMoyano2009}). 
{\it Uniform} means that all the $W$ windows start with the same number 
of initial conditions; 
{\it alternate} means that the occupancy of each of the $W$ windows starts 
alternating the double of initial conditions and emptiness, 
the total number $WM$ of initial conditions in the interval 
$[-1,1]$ remaining the same.
}
\label{phasespace}
\end{figure}

\begin{figure}[h!]
\centering
\includegraphics[scale=0.45,angle=0]{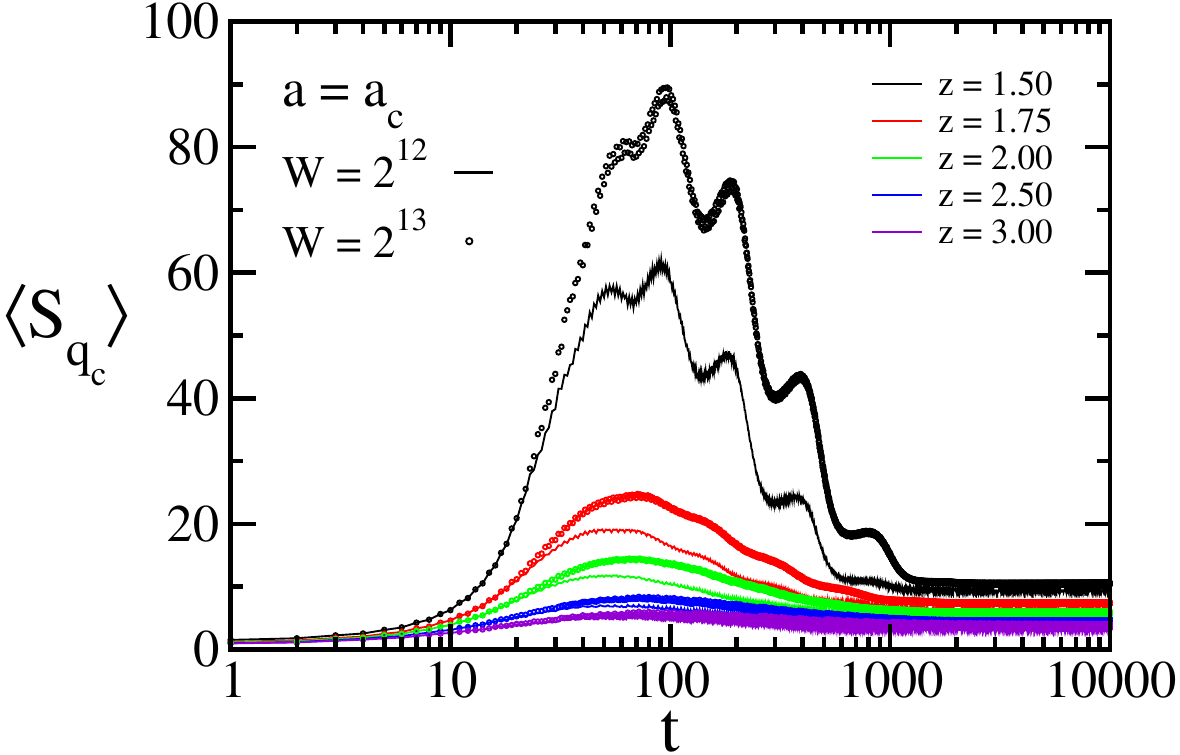}
\caption{\small (Color online) 
Peaks corresponding to the $z$-logistic map $x_{t+1}=1-a|x_t|^z$ 
for typical values of $z$. 
For $z=1.50,1.75,2.00,2.50,3.00$ we have 
$(a_c,q_c)= (1.295509973160,  -0.15)$,
$(1.355060756622,   0.11)$,
$(1.40115518909205, 0.2444877)$,
$(1.47054991523,    0.39)$,
$(1.52187878890,    0.47)$.
}
\label{zmaps}
\end{figure}

Before concluding, let us emphasize that the robustness of the peaks 
in the time evolution of the entropy $S_q$ at $a=a_c$ 
has been numerically verified under four different circumstances, 
namely with regard to 
(i)   the number of initial conditions within each one of the $W$ windows 
      (Fig. \ref{initialconditions}); 
(ii)  the precision used in the value of $a_c$ 
      (bottom-right plot in Fig. \ref{fig3}); 
(iii) variations of the occupancy of phase space in the neighborhood 
      of the multifractal attractor at $a=a_c$ 
      (Fig. \ref{phasespace}); 
(iv)  variations of the map (Fig. \ref{zmaps}).

\subsection{4 - Final remarks}

At this point, a few comments are certainly timely concerning the most 
distinguished non-trace-form entropic functional, 
namely the Renyi one \cite{Renyi1961} 
\begin{equation}
S_q^R \equiv k\frac{\ln \sum_{i=1}^Wp_i^q}{1-q}=k\frac{\ln[1+(1-q)S_q/k]}{1-q} 
\;\;(q \in \mathbb{R}; \,S_1^R=S_1=S_{BG})\,,
\end{equation} 
hence
\begin{equation}
S_q^R/k=\ln e_q^{S_q/k}
\end{equation}
and
\begin{equation}
S_q/k=\ln_q e^{S_q^R/k}
\end{equation}
This additive (hence composable) entropic functional is, $\forall q$, 
a monotonic function of $S_q$, hence, under the same constraints, 
it is optimized by the same distribution which optimizes $S_q$. 
However, in contrast with $S_q$ which is concave for all $q>0$, 
$S_q^R$ is concave only for $0<q \le 1$. 
In addition to that, $S_q$ is Lesche-stable for all $q>0$, 
whereas $S_q^R$ has this important experimental robustness 
only for its particular instance $q=1$ 
\cite{Lesche1982,Abe2002,AbeLescheMund2007}. 
Moreover, if we consider the present nontrivial case $a=a_c$, 
we have that, in the $W\to\infty$ limit, 
$S_{q_c}^R(t)=k\frac{\ln[1+(1-q_c)S_{q_c}(t)/k]}{1-q_c} \propto \ln t$ 
for $t \gg 1$. 
Consequently, there is no constant finite Renyi-entropy production 
per unit time and, consistently, no Pesin-like identities can exist. 
Analogously, if we consider the stationary-state of a $N$-body system 
belonging to the previously mentioned anomalous class 
$W(N) \propto N^\rho$, we have that $S_q^R(N) \propto \ln N$ for any $q<1$. 
In other words, $S_q^R(N)$ appears to be nonextensive 
and the Legendre structure of thermodynamics is therefore violated. 

It is worthy to mention that composability is not necessary for a finite 
entropy production per unit time to exist 
(thus possibly qualifying for a Pesin-like identity to be satisfied). 
Such is the case, for instance, of the Kaniadakis entropy \cite{Kaniadakis2001} 
which is nonadditive, non-composable, trace-form, and yields nevertheless 
a finite entropy production per unit time.

\begin{figure}[h!]
\centering
\includegraphics[scale=0.45,angle=0]{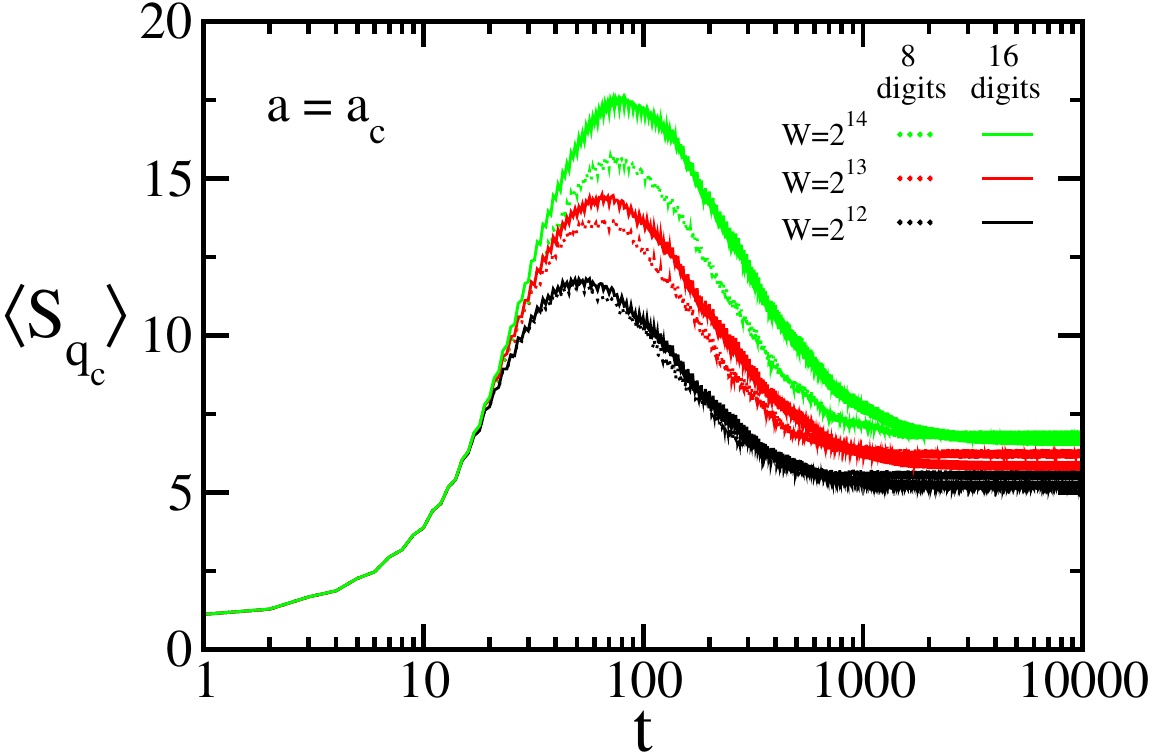}
\includegraphics[scale=0.45,angle=0]{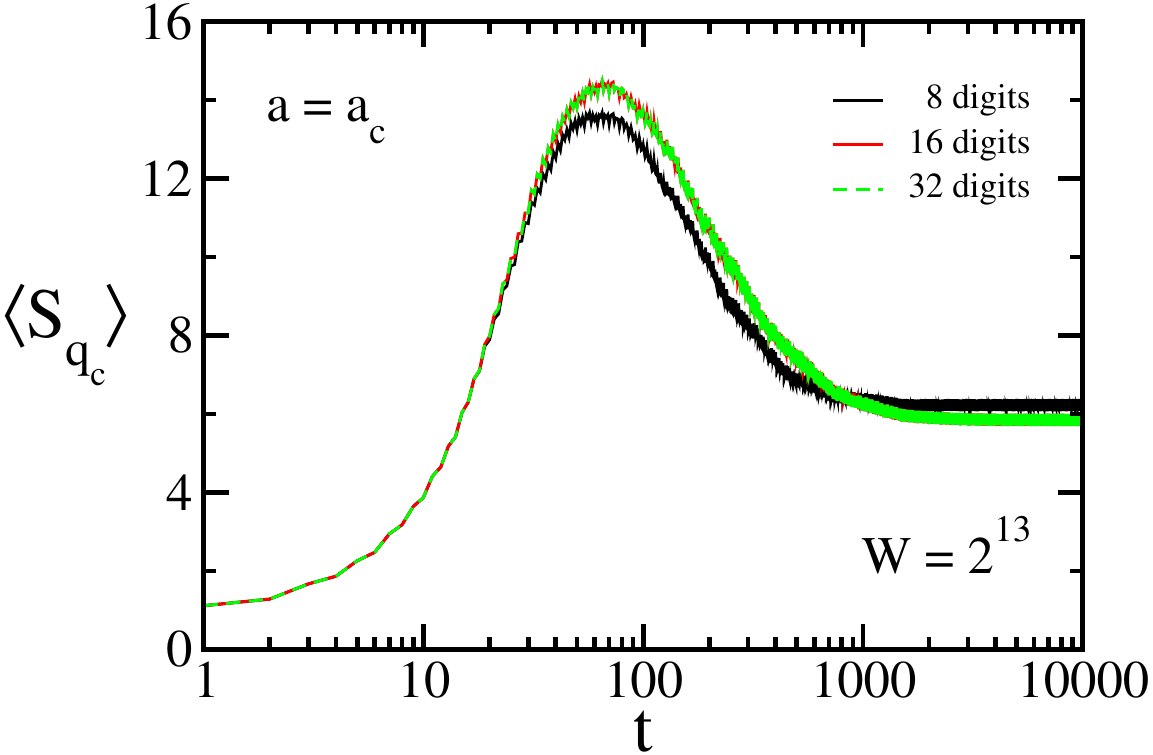}
\includegraphics[scale=0.45,angle=0]{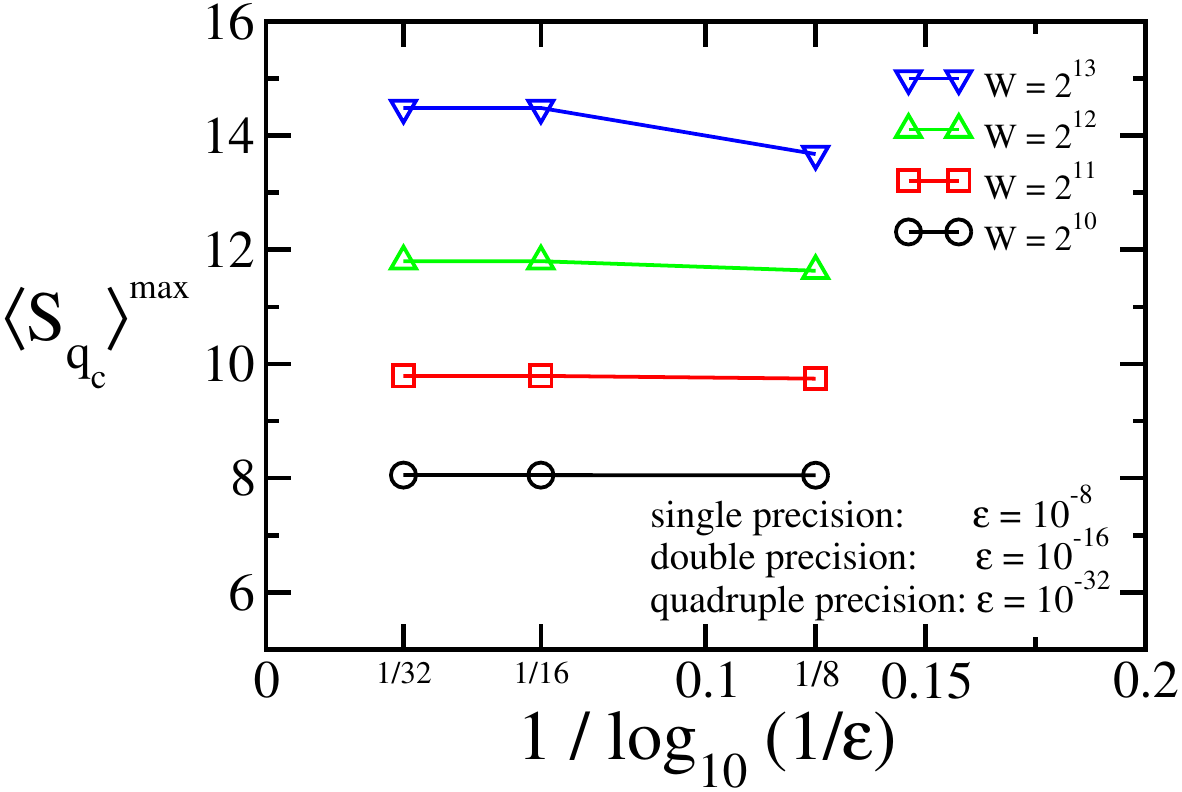}
\caption{\small (Color online) 
Influence of the precision of the calculations. 
We verify that, for small values of $W$, double precision (i.e., 16 digits) 
is more than enough for the present purposes, but, when $W$ sensibly increases, 
increasingly higher precision becomes necessary. 
Interestingly enough, low precision underestimates the maximal value of 
$\langle S_{q_c}\rangle$, whereas it overestimates its stationary-state value. 
}
\label{fig6}
\end{figure}

The relevant influence of the precision used in the calculations 
has been exhibited as well: see Fig.\ \ref{fig6}. 
More specifically, the use of simple, double or quadruple precision sensibly 
enhances the overshooting of the time evolution of the entropy $S_{q_c}(t)$ 
at the Feigenbaum point. 
To reinforce these results, it would of course  be necessary to use values of 
$a_c$ and of $q_c$ with consistently higher precision when $W$ increases. 
Still, the present indications are already clear enough.

Let us conclude by focusing on a relevant result of the present study. 
Our numerical simulations strongly indicate that, for both strong and weak 
chaos, the time behavior of $S_q$ 
($q=1$ for $a=2$ and $q=q_c$ for $a=a_c$, respectively) 
at the $W \to\infty$ limit, consists in a diverging linear increase  
with finite slope. 
However, an important distinction arises on how $S_q(t)$ approaches 
its stationary-state value $S_q(\infty)$: 
it approaches {\it from below} for $a=2$, which corresponds to the naive 
expectation, whereas it does so {\it from above} for $a=a_c$, 
which might be considered as unexpected. 
This -- a priori surprising -- behavior is due to the fact that, 
for relatively early times during the present evolution at fixed $W$, 
the entropy system tends to approach its maximal value $\ln_q W$. 
However, for later times, the system approaches its asymptotic stationary state,
{\it which is not uniform} 
(it is instead $U$-shaped for $a=2$ and multifractal for $a=a_c$). 

The 2nd principle of thermodynamics is normally qualified for systems 
which are closed, very large, and for a generic initial situation. 
An intriguing question might arise: these conditions surely are necessary, 
but are they sufficient? 
Is there no need for also requiring a strongly chaotic internal dynamics 
(e.g., short-range interactions), which normally implies mixing and ergodicity 
for all or part of the system? 
At the light of the present results, this fundamental question appears to be 
an open one, surely deserving further study.

%\section*{Acknowledgments}
We acknowledge fruitful conversations with E.M.F. Curado 
and partial financial support by CNPq and Faperj (Brazilian agencies). 
We also acknowledge useful remarks from two anonymous referees 
which led to various numerical verifications of the robustness of the entropic 
peaks at the Feigenbaum point, thus enriching the manuscript.

\end{document}